\documentstyle[prl,aps,prb]{revtex}

\begin{document}

\thispagestyle{empty}

\begin{center}
\vskip100pt
{ \Large {\bf
ANISOTROPY OF FAST-GOING PROCESSES IN THE SUN
\vskip8pt
AND NEW INTERACTION IN NATURE
}}
\vskip20pt
Yu.A. Baurov \vskip15pt

{\it Central Scientific Research Institute of Machinery

 141070, Kaliningrad, Moscow region, Russia}
\vskip20pt

A.A. Efimov \&  A.A. Shpitalnaya\vskip15pt

{\it State Astronomical Observatory of Russian Academy of Sciences,

190000, Pulkovo, Leningrad region, Russia
}
\vskip30pt
ABSTRACT
\end{center}

The processing of the galactic coordinates of 3543 solar flares
with a magnitude of 2 or more has shown  that the distribution
of these fast-going processes on the surface of the "nonrotating" Sun
is irregular and non-random what testifies that in the near-Sun space
an anisotropy takes place which practically  coincides with that predicted
and obtained in laboratory experiments and caused by existence of
the intergalactic vector potential ${\bf A_g}$.
\vskip5pt
Subject headings: Sun: flares - Sun: magnetic fields - Cosmology: theory.
\pagebreak
\section{Introduction and motivation.}

In the works $^{1-5}$ the results of experimental investigations to find
a new natural interaction arising when magnet systems act on physical vacuum
 through their vector  potential, are presented. The essence of the
 new interaction lies in the fact that according to developed in Refs. $^{6,7}$
 ideas of physical vacuum structure, masses of elementary particles
are proportional to the magnitude of the intergalactic vector potential
${\bf A_g}$, a new fundamental  vectorial constant, related to one-dimensional
discrete "magnetic" fluxes which, by the model of the Universe $^{8,9}$,
 form our entire world. In the theory developed, processes of forming
elementary particle charge numbers are investigated (in particular, with
 variation of electric charge and, hence, violation of gauge invariance),
 therefore potentials, at such an approach, assume a physical meaning.
 By the theory the modulus $\vert {\bf A_g}\vert$ has a limiting value
$\vert {\bf A_g}\vert \approx 1,95\times10^{11}$ CGSE
 units)  and cannot be increased but diminished, for instance, by the vector
 potential of a certain magnet directed towards ${\bf A_g}$.  Inasmuch as masses of
elementary particles are uniquely related to the value $\vert {\bf A_g}\vert^{6,7}$, an
 assumption may be made about existence, in a region with lowered
$\vert {\bf A_g}\vert$, of a new type of interaction, acting on any material body located there.

The new force is essentially non-linear and non-local in character with
 respect to the value $\Delta A$, a variation of the modulus of ${\bf A_g}$ , and to the gradient
of $\Delta A$ in direction of the vector  ${\bf A_g}$$^{4,9}$. This force is mainly in the same
 direction as the vector ${\bf A_g}$  which according to Refs.$^{1-5}$ has the following
 coordinates: right ascension $\alpha\approx 270^\circ\pm 7^\circ$,
declination $\delta\approx +30^\circ$ (the second
equatorial coordinate system).

Thus, the direction of the vector ${\bf A_g}$  determines global space anisotropy
in the vicinity of the Sun. It is necessary to note that the measured
 direction of ${\bf A_g}$  under terrestrial conditions can differ from the true
one because the fundamental vector ${\bf A_g}$  is superimposed by vector potentials
of galactic and intergalactic magnetic fields, i.e. on the Earth (in the
 Sun's vicinity) we measure, really, a summary potential  ${\bf A_\Sigma}$ which can be
some what lesser in magnitude than $\vert {\bf A_g} \vert$ and rotated relative to the true
undisturbed vector  ${\bf A_g}$ . The new interaction has to show itself not only
in the process of interacting of the entire Sun's magnetic system with
 physical vacuum $^{5,9}$ but also in phenomena associated with fast-going
 electromagnetic processes in the Sun proceeding with energy release.
 Among these are solar flares and a spontaneous disappearing of fibres,
 solar eruptive prominences $^{10-12}$. Here we shall demonstrate that the
 solar flares with a magnitude of 2 or more are distributed over the surface
 of the "stationary" Sun anisotropically with the maximum of this anisotropy
 practically coinciding with the direction of the vector  ${\bf A_g}$.

\section{The new interaction and solar flares.}

Let us refer to Fig.1 where is shown a diagram of the new force $\overrightarrow{F}$
($\overrightarrow{\bf A_g}\|\overrightarrow{F}$) action
 on a magnetic tube (1) located under the Sun's  surface (2) and related
to sunspots (3). As a rule, the flares are observed in large groups of
 sunspots with a complex configuration of magnetic field and are determined,
 in accordance with the available models $^{12}$, by storing magnetic energy in
 the upper chromosphere and the lower corona. The crosshatched area shows
 schematically a space region with the vector potential $\overrightarrow{A}$ of the tube (1)
 directed towards the vector ${\bf A_g}$. Just in this region
the new force $\overrightarrow{F}$ arises,
 which forms, in our opinion, the space anisotropy as the result of Archimedes'
 floating up of the tubes.

The new interaction must be very strong in that case because the vector
 potential $\overrightarrow{A}$ of the tubes is of the order of
$\sim 10^{11}$ Gs$\cdot$cm and comparable to the value $\vert {\bf A_g}\vert$
( the mean diameter of the flare is equal to $5\times10^9$cm,
 the magnetic field value $B\approx 50$ Gs; the typical sunspot diameter
is about $10^8$ cm, the greatest value of $B$ goes up to $5000$ Gs;
in Ref.$^{13}$ it is shown that the splitting  of  metal  spectral lines observed
in solar flares requires  that the fields  be about $10^4$ Gs).
These circumstances lead, in our opinion, to the observed anisotropy
 in solar flare distribution on the surface of the "stationary" Sun.

\section{Method of investigation and results.}

In Refs.$^{14,15}$, as a result  of investigating distribution of a large number
 of different nonstationary processes in our Galaxy including also solar flares
 (in galactic longitude), the clearly defined space anisotropy was found.
In the work $^{16}$ proposed, and in $^{17}$ realized is a differential method
 of determining the direction of the maximum space anisotropy.

Processing coordinates of 3324 solar flares with a magnitude 2 or more taken
 from catalogues $^{18-21}$ has given the following results in the second
 equatorial coordinate system: $\alpha\approx 271^\circ, \delta\approx + 15^\circ$.

We proposed an integral method of determining the direction to be found.
For this purpose the entire Sun's surface was divided into 648 elements
 of area measuring $10^\circ_l\times10^\circ_b$ according to 648 directions, onto which the unit vectors
 were projected whose ends corresponded to the spherical coordinates of 3543
 solar flares with magnitude $\ge 2$ (the galactic coordinate system was used:
 $l$ is longitude, $b$ is latitude). The sum of all the projections of the unit
 vectors onto a choiced direction $i$ ( $\sum\limits_\gamma \cos C_\gamma$ , where $\gamma$  is an index characterizing
 the spherical coordinates of a flare) served as a characteristic of $i$-th
 semisphere. In Fig.2 for semispheres situated in the same diameter
coinciding with the $i$-th direction, a matrix of values $\alpha_i$, the deviations
 of relative frequency from the most probable value $\sigma$ (the isotropic
 distribution), which matrix was computated by the method proposed
in Ref.$^{22}$, is presented.

At a concrete value of the index $i$  we have for $\alpha$:
$$ \frac{\alpha}{\sigma} = \frac{q\Sigma_1 - p\Sigma_2}{\Sigma_1 + \Sigma_2}
\sqrt{\frac{\Sigma_1 + \Sigma_2}{qp}},\eqno(1)$$
where $\Sigma_1 = ( \sum\limits_\gamma \cos C_\gamma )_{up}$ denotes the magnitude of the sum of cosines for the semisphere
 from which vector $i$ comes out; $\Sigma_2 = ( \sum\limits_\gamma \cos C_\gamma )_{down}$
  is an analogous demotion for the diametrically opposed semisphere;
$p$ and $q$ are probabilities of occurrence of $\Sigma_1$  and $\Sigma_2$  values.

In our case $p = q = \frac{1}{2}$, i.e. the formula (1) may be rearranged to
$$ \alpha = \frac{\Sigma_1 - \Sigma_2}{\sqrt{\Sigma_1 + \Sigma_2}}\sigma \eqno(2)$$
In the absence of anisotropy the equality of sums $\Sigma_1 = \Sigma_2$ would take place,
 and the isotropic distribution would exist. In our case for the extremal
 anisotropy we have $\Sigma_1 = 878,\; \Sigma_2 = 555$, i.e. $\alpha = 8.5\sigma$. It is seen from Fig.2
 that the direction of maximum anisotropy for the diametrically opposite
 Sun's semispheres lies in the region of standard Sun's apex and has
 the following coordinates:
$$   \max \left\{\alpha (\hbox{ right ascension} ) \approx 277^\circ\atop\delta (\hbox{ declination} ) \approx + 38^\circ\right\}       $$
(the second equatorial
 coordinate system), $ l = 65^\circ,\; b = + 20^\circ$ (the galactic coordinate system).
The methodical error is equal to $\pm 5^\circ$.

The found coordinates practically coincide with the experimentally determined
 coordinates of the vector ${\bf A_g}$$^{1-5}$ and are trustworthy since the observed
 anisotropy accounts for more than $8\sigma$  with respect to the isotropic state.
Thus, the space anisotropy associated with the vector ${\bf A_g}$ plays, as is seen,
 an important part in nonstationary explosive processes of electromagnetic
nature in the solar plasma.



Legends of the figures :

Fig.1.  A schematic sketch of acting of the new force on a magnetic tube.

1 - magnetic tube,      2 - Sun's surface,        3 - sunspot,

4 - direction of the magnetic field  in a tube;

$\overrightarrow A$ - direction of vector potential of the magnetic field $ B$;

${\bf A_g}$ - direction of the intergalactic vector potential;

5 - region of lowered magnitude of the vector ;

$\overrightarrow F$ - direction of the new force;  Z - zenith point.

\newpage
\textwidth  200mm
\oddsidemargin -.7in
\begin{center}

Fig.2.  Magnitude of the quantity $\alpha$  in $i$-th direction in the galactic coordinate system.

\footnotesize

\vskip10pt

\begin{tabular}{c|c|c|c|c|c|c|c|c|c|c|c|c|c|c|c|c|c|c|}
\hline

\hss$ l\backslash b$\hss &$85^\circ$&$75^\circ$&$65^\circ$&$55^\circ$&$45^\circ$&$35^\circ$&$25^\circ$&$15^\circ$&$5^\circ$&$-5^\circ$&$-15^\circ$&$-25^\circ$&$-35^\circ$&$-45^\circ$&$-55^\circ$&$-65^\circ$&$-75^\circ$&$-85^\circ$\\
\hline
\hss$0^\circ$\hss &2.6&3.5 &4.2 &4.8&5.2&5.5&5.7&5.6&5.4&5.1&4.5&3.9&3.1&2.3&1.3&0.3&-0.7&-1.7\\
\hline

\hss$10^\circ$\hss &2.7&3.7&4.6&5.3&5.8&6.3&6.4&6.4&6.2&5.8&5.2&4.5&3.7&2.8&1.7&0.6&-0.5&-1.6\\
\hline

\hss$20^\circ$\hss &2.8&3.9&4.9&5.7&6.4&6.9&7.2&6.9&6.8&6.4&5.8&5.0&4.1&3.1&2.0&0.9&-0.3&-1.6\\
\hline

\hss$30^\circ$\hss &2.8&4.0&5.1&6.1&6.9&7.4&7.7&7.6&7.4&6.9&6.2&5.4&4.4&3.4&2.2&1.0&-0.2&-1.5\\
\hline

\hss$40^\circ$\hss &2.8&4.1&5.3&6.4&7.2&7.8&{\bf 8.2}&{\bf 8.1}&7.8&7.2&6.4&5.5&4.6&3.5&2.3&1.1&-0.2&-1.5\\
\hline

\hss$50^\circ$\hss &2.8&4.1&5.4&6.5&7.5&{\bf 8.2}&{\bf 8.6}&{\bf 8.5}&{\bf 8.1}&7.5&6.6&5.6&4.6&3.4&2.2&1.0&-0.2&-1.5\\
\hline

\hss$60^\circ$\hss &2.8&4.0&5.3&6.5&7.6&{\bf 8.3}&{\bf 8.9}&{\bf 8.8}&{\bf 8.3}&7.5&6.6&5.5&4.4&3.2&2.1&0.9&-0.3&-1.5\\
\hline

\hss$70^\circ$\hss &2.8&4.0&5.2&6.4&7.6&{\bf 8.5}&{\bf 8.9}&{\bf 8.8}&{\bf 8.3}&7.3&6.3&5.1&4.0&2.9&1.8&0.7&-0.4&-1.6\\
\hline

\hss$80^\circ$\hss &2.7&3.8&4.9&6.1&7.2&{\bf 8.0}&{\bf 8.3}&{\bf 8.3}&7.8&6.8&5.7&4.6&3.5&2.4&1.4&0.4&-0.6&-1.6\\
\hline

\hss$90^\circ$\hss &2.6&3.5&4.5&5.6&6.5&7.2&7.3&7.2&6.8&5.8&4.8&3.7&2.8&1.8&0.9&0.0&-0.8&-1.7\\
\hline

\hss$100^\circ$\hss &2.5&3.3&4.0&4.8&5.6&5.8&6.1&5.9&5.4&4.5&3.6&2.7&1.9&1.1&0.4&-0.4&-1.1&-1.8\\
\hline

\hss$110^\circ$\hss &2.4&2.9&3.5&3.9&4.4&4.5&4.6&4.3&3.7&3.0&2.3&1.6&1.0&0.3&-0.2&-0.8&-1.4&-1.9\\
\hline

\hss$120^\circ$\hss &2.3&2.6&2.8&3.1&3.2&3.1&2.9&2.5&2.0&1.4&0.9&0.4&-0.1&-0.5&-0.9&-1.3&-1.7&-2.0\\
\hline

\hss$130^\circ$\hss &2.2&2.3&2.2&2.1&2.0&1.7&1.3&0.9&0.4&0.0&-0.2&-0.8&-1.1&-1.4&-1.6&-1.8&-2.0&-2.1\\
\hline

\hss$140^\circ$\hss &2.1&1.9&1.6&1.3&0.9&0.4&-0.1&-0.5&-1.0&-1.4&-1.7&-1.9&-2.1&-2.2&-2.3&-2.3&-2.3&-2.2\\
\hline

\hss$150^\circ$\hss &2.0&1.5&1.0&0.5&-0.1&-0.7&-1.3&-1.8&-2.2&-2.6&-2.8&-3.0&-3.1&-3.1&-3.0&-2.8&-2.6&-2.3\\
\hline

\hss$160^\circ$\hss &1.9&1.2&0.5&-0.2&-1.0&-1.7&-2.3&-2.8&-3.3&-3.6&-3.9&-4.0&-4.0&-3.9&-3.6&-3.3&-2.9&-2.4\\
\hline

\hss$170^\circ$\hss &1.8&0.9&0.0&-0.8&-1.7&-2.5&-3.2&-3.8&-4.2&-4.6&-4.8&-4.9&-4.8&-4.6&-4.2&-3.8&-3.2&-2.5\\
\hline

\hss$180^\circ$\hss &1.7&0.7&-0.3&-1.3&-2.3&-3.1&-3.9&-4.5&-5.1&-5.4&-5.6&-5.7&-5.5&-5.2&-4.8&-4.2&-3.5&-2.6\\
\hline

\hss$190^\circ$\hss &1.6&0.5&-0.6&-1.7&-2.8&-3.7&-4.5&-5.2&-5.8&-6.2&-6.4&-6.4&-6.3&-5.8&-4.8&-4.6&-3.7&-2.7\\
\hline

\hss$200^\circ$\hss &1.6&0.3&-0.9&-2.0&-3.1&-4.1&-5.0&-5.8&-6.4&-6.8&-6.9&-7.2&-6.9&-6.4&-5.3&-4.9&-3.9&-2.8\\
\hline

\hss$210^\circ$\hss &1.5&0.2&-1.0&-2.2&-3.4&-4.4&-5.4&-6.2&-6.9&-7.4&-7.6&-7.7&-7.4&-6.9&-5.7&-5.1&-4.0&-2.8\\
\hline

\hss$220^\circ$\hss &1.5&0.2&-1.1&-2.3&-3.5&-4.6&-5.5&-6.4&-7.2&-7.8&-{\bf 8.1}&-{\bf 8.2}&-7.8&-7.2&-6.1&-5.3&-4.1&-2.8\\
\hline

\hss$230^\circ$\hss &1.5&0.2&-1.0&-2.2&-3.4&-4.6&-5.6&-6.6&-7.5&-{\bf 8.1}&-{\bf 8.5}&-{\bf 8.6}&-{\bf 8.2}&-7.5&-6.4&-5.4&-4.1&-2.8\\
\hline

\hss$240^\circ$\hss &1.5&0.3&-0.9&-2.1&-3.2&-4.4&-5.5&-6.6&-7.5&-{\bf 8.3}&-{\bf 8.8}&-{\bf 8.9}&-{\bf 8.3}&-7.6&-6.5&-5.3&-4.0&-2.8\\
\hline

\hss$250^\circ$\hss &1.6&0.4&-0.7&-1.8&-2.9&-4.0&-5.1&-6.3&-7.3&-{\bf 8.3}&-{\bf 8.8}&-{\bf 8.9}&-{\bf 8.5}&-7.6&-6.5&-5.2&-4.0&-2.8\\
\hline

\hss$260^\circ$\hss &1.6&0.6&-0.4&-1.4&-2.4&-3.5&-4.6&-5.7&-6.8&-7.8&-{\bf 8.3}&-{\bf 8.3}&-{\bf 8.0}&-7.2&-6.4&-4.9&-3.8&-2.7\\
\hline

\hss$270^\circ$\hss &1.7&0.8&0.0&-0.9&-1.8&-2.8&-3.7&4.8&-5.8&-6.8&-7.2&-7.3&-7.2&-6.5&-6.1&-4.5&-3.5&-2.6\\
\hline

\hss$280^\circ$\hss &1.8&1.1&0.4&-0.4&-1.1&-1.9&-2.7&-3.6&-4.5&-5.4&-5.9&-6.1&-5.8&-5.6&-5.6&-4.0&-3.3&-2.5\\
\hline

\hss$290^\circ$\hss &1.9&1.4&0.8&0.2&-0.3&-1.0&-1.6&-2.3&-3.0&-3.7&-4.3&-4.6&-4.5&-4.4&-4.8&-3.5&-2.9&-2.4\\
\hline

\hss$300^\circ$\hss &2.0&1.7&1.3&0.9&0.5&0.1&-0.4&-0.9&-1.4&-2.0&-2.5&-2.9&-3.1&-3.2&-3.9&-2.8&-2.6&-2.3\\
\hline

\hss$310^\circ$\hss &2.1&2.0&1.8&1.6&1.4&1.1&0.8&0.2&0.0&-0.4&-0.9&-1.3&-1.7&-2.0&-3.1&-2.2&-2.3&-2.2\\
\hline

\hss$320^\circ$\hss &2.2&2.3&2.3&2.3&2.2&2.1&1.9&1.7&1.4&1.0&0.5&0.1&-0.4&-0.9&-2.1&-1.6&-1.9&-2.1\\
\hline

\hss$330^\circ$\hss &2.3&2.6&2.8&3.0&3.1&3.1&3.0&2.8&2.6&2.2&1.8&1.3&0.4&0.1&-1.3&-1.0&-1.5&-2.0\\
\hline

\hss$340^\circ$\hss &2.4&2.9&3.3&3.6&3.9&4.0&4.0&3.9&3.6&3.3&2.8&2.3&1.7&1.0&-0.5&-0.5&-1.2&-1.9\\
\hline

\hss$350^\circ$\hss &2.5&3.2&3.8&4.2&4.6&4.8&4.9&4.8&4.6&4.2&3.8&3.2&2.5&1.7&0.2&0.0&-0.9&-1.8\\
\hline

\hss$360^\circ$\hss &2.6&3.5&4.2&4.8&5.2&5.5&5.7&5.6&5.4&5.1&4.5&3.9&3.1&2.3&1.3&0.3&-0.7&-1.7\\
\hline

\end{tabular}

\end{center}


\begin{thebibliography}{confr}

\bibitem{1}
Baurov Yu.A., Klimenko ….Yu. \& Novikov S.I. 1990, Docl. Akad. Nauk SSSR, 315, 1116

\bibitem{2}
Baurov Yu.A., Klimenko ….Yu. \& Novikov S.I. 1992, Phys. Lett., {\bf A162}, 32

\bibitem{3}
Baurov Yu.A. \& Rjabov P.Œ. 1992, Docl. Akad. Nauk SSSR, 326, 73

\bibitem{4}
Baurov Yu.A. 1993, Phys. Lett., {\bf A181}, 283

\bibitem{5}
Baurov Yu.A., Serjogin B.Œ. \& Chernikov A.V. 1994, Fiz. Mysl Ross., N1, 1

\bibitem{6}
Babajev Yu.N. \& Baurov Yu.A. 1984, preprint {\bf -0362} (Moskow: INR Akad. Nauk SSSR)

\bibitem{7}
Babajev Yu.N. \& Baurov Yu.A. 1985, preprint {\bf -0386} (Moskow: INR Akad. Nauk SSSR)

\bibitem{8}
Baurov Yu.A. 1990, in:  Plasma  physics  and  certain  problems  of  general  physics
(Kaliningrad, Moscow reg.:  Central Scientific Research Institute of Machinery ), pp. 71, 84, [in Russian]

\bibitem{9}
Baurov Yu.A. 1994, Fiz. Mysl Ross., N1, 18

\bibitem{10}
Zirin G. 1969, Solar atmosphere (Moskow: "Mir"), [in Russian]

\bibitem{11}
ed. Coyper J. 1957, Sun  System.  Sun. (Moskow : Inosrtannaya Literatura), v.1, [in Russian]

\bibitem{12}
Somov B.V., Syrovatsky S.I. 1978, Usp. Fiz. Nauk, 120, 217

\bibitem{13}
Alikaeva Š.V. 1969, Astrometric and Astrophysics, 8, 92

\bibitem{14}
Shpitalnaya A.A. 1978, Probl. Issl. Vsel., 7, 193

\bibitem{15}
Shpitalnaya A.A. 1979, Probl. Issl. Vsel., 8, 264

\bibitem{16}
Efimov A.A. 1979, Probl. Issl. Vsel., 8, 567

\bibitem{17}
Efimov A.A.\& Shpitalnaya A.A. 1980, Probl. Issl. Vsel., 9, 67

\bibitem{18}
Fritzova L., Kopecky M., Svestka Z. 1958,  J. Astr. Inst., Catalogue of Great Chromospheric Flares and their Terrestrial Consequences (SSSR), 35

\bibitem{19}
Catalogue of Solar Particle Events 1955-1969. 1970, (Dordrecht-Holland, Boston USA)

\bibitem{20}
Quarterly Bulletin on Solar Activity 1939-1980. 1982, (Z\"urich)

\bibitem{21}
Dolguinova Yu.N. 1972, Solar-Terrestrial Physics. Catalogue of Chromospheric Flares (Moskow), 2, [in Russian]

\bibitem{22}
Agekjan T.A. 1972, Theory of Errors (Moskow), p.75, [in Russian]

\end{thebibliography}
\end{document}